\documentclass[conference]{IEEEtran}
\IEEEoverridecommandlockouts
\usepackage{cite}
\usepackage{amsmath,amssymb,amsfonts}
\usepackage{algorithm}
\usepackage[noend]{algpseudocode}
\usepackage{graphicx}
\usepackage{textcomp}
\usepackage{nccmath}
\usepackage{xcolor}
\usepackage{lscape}
\usepackage{mathtools}
\usepackage{stackrel}
\usepackage{times}
\usepackage{latexsym}
\usepackage{textcomp}
\usepackage{xcolor}
\usepackage{dirtytalk}
\usepackage{caption}
\usepackage{array,multirow}
\usepackage{float}
\usepackage{hyperref}
\usepackage{mathtools}
\usepackage{textcomp}
\usepackage{dirtytalk}
\usepackage{latexsym}
\usepackage{lineno,hyperref}
\usepackage{bbm}
\usepackage{booktabs} % For formal tables
\usepackage{tabularx}

\usepackage{algpseudocode}
\def\BibTeX{{\rm B\kern-.05em{\sc i\kern-.025em b}\kern-.08em
T\kern-.1667em\lower.7ex\hbox{E}\kern-.125emX}}
\begin{document}

\title{A Personalized Reinforcement Learning Summarization Service for\\Learning Structure from Unstructured Data}

% \author{\IEEEauthorblockN{1\textsuperscript{st} Samira Ghodratnama}
% \IEEEauthorblockA{\textit{{Computing Department, Grainger Technology Group}} \\
% \textit{Macquarie Univeristy, Grainger}\\
% {Sydney Australia, Chicago USA}\\
% samira.ghodratnama@mq.edu.au,\\samira.ghodratnama@grainger.com}
% \and
% \IEEEauthorblockN{2\textsuperscript{nd} Amin Beheshti}
% \IEEEauthorblockA{\textit{Computing Department} \\
% \textit{Macquarie University}\\
% Sydney, Australia \\
% amin.beheshti@mq.edu.au}
% \and
% \IEEEauthorblockN{3\textsuperscript{rd} Mehrdad Zakershahrak}
% \IEEEauthorblockA{\textit{Computing Department} \\
% \textit{Macquarie University}\\
% Sydney, Australia \\
% mehrdad.zakershahrak@mq.edu.au}
% }

\author{
  \IEEEauthorblockN{Samira Ghodratnama}
  \IEEEauthorblockA{Macquarie University, Australia}
  \IEEEauthorblockA{W.W. Grainger, USA} 
  \IEEEauthorblockA{samira.ghodratnama@mq.edu.au}
  \IEEEauthorblockA{samira.ghodratnama@grainger.com}
  \and
  
  \IEEEauthorblockN{Amin Behehsti}
  \IEEEauthorblockA{Macquarie University, Australia} 
  \IEEEauthorblockA{amin.beheshti@mq.edu.au}
  \and
  
  \IEEEauthorblockN{Mehrdad Zakershahrak}
  \IEEEauthorblockA{Macquarie University, Australia} 
  \IEEEauthorblockA{mehrdad.zakershahrak@mq.edu.au}

}

\maketitle

\begin{abstract}
The exponential growth of textual data has created a crucial need for tools that assist users in extracting meaningful insights. Traditional document summarization approaches often fail to meet individual user requirements and lack structure for efficient information processing. To address these limitations, we propose Summation, a hierarchical personalized concept-based summarization approach. It synthesizes documents into a concise hierarchical concept map and actively engages users by learning and adapting to their preferences. Using a Reinforcement Learning algorithm, Summation generates personalized summaries for unseen documents on specific topics. This framework enhances comprehension, enables effective navigation, and empowers users to extract meaningful insights from large document collections aligned with their unique requirements.
\end{abstract}


\section{Introduction} \label{intro}
The availability of a vast amount of information on various topics has led to a phenomenon known as \textit{information overload}, where the volume of data exceeds an individual's capacity for effective processing within a reasonable timeframe.
While this abundance of data can be valuable for analytical applications, it necessitates efficient exploration tools to harness its potential benefits without succumbing to information overload, which can strain cognitive resources.
Data summaries serve as effective tools for gathering relevant information, organizing it into a coherent and manageable form, and facilitating complex question answering, insight generation, and conceptual boundary discovery~\cite{schiliro2019icop,amouzgar2019isheets,beheshti2018iprocess}.
Automatic document summarization has been extensively studied to address the challenges of data reduction for analysis, commercialization, management, and personalization purposes.
Furthermore, users often seek information in an organized and coherent structure.
However, despite the speed of document generation and the massive collections of unstructured documents, producing personalized summaries comparable to human-written ones remains challenging.
Most previous work on automatic text summarization has focused on generating textual summaries rather than structured ones.
These approaches typically produce a single, short, general, and flat summary that applies to all users, lacking interpretability and personalization. 
Moreover, they are incapable of producing more extended and detailed summaries, even if users express interest in obtaining additional information.
Additionally, the lack of structure in these summaries hampers further processing, and they heavily rely on reference or gold summaries created by humans, which are subjective and costly \cite{ghodratnama2021intelligent,khanna2022transformer}. 
To address these limitations, we propose \textit{Summation}, a hierarchically interactive structured summarization approach that generates personalized summaries. 
We emphasize the significance of the following aspects in our contribution: i) Structured summaries, ii) Personalization, iii) Interaction, and iv) The elimination of reference summaries.  

\textbf{Structured Summaries.} Studies have demonstrated that when individuals encounter numerous documents, they seldom formulate fully-fledged summaries. Instead, they attempt to extract concepts and understand the relationships among them~\cite{beheshti2022social,gupta2010survey,beheshti2021bp}. 
Consequently, structured data has become crucial in various domains.
It offers a concise overview of the document collection's contents, unveils interesting relationships, and serves as a navigational structure for further exploration of the documents. 
Our approach, \textit{Summation}, provides summaries in the form of a hierarchical concept map, which caters to diverse user requirements by being interpretable, concise, and simultaneously providing an overview and detailed information. 

\textbf{Personalization.} Existing summarization approaches typically generate a generic summary comprising a few selected sentences intended to meet the needs of all users. In contrast to such generic summaries, there is a dearth of user-centric summarization approaches that allow users to specify the desired content in the summaries~\cite{ghodratnama2021towards,ghodratnama2021summary2vec}.

\textbf{Interaction.} Conventional summarization approaches treat a topic-related document set as input and generate a summary that captures the most salient aspects. However, research on this topic often neglects the usefulness of the approach for users, focusing primarily on the accuracy of the generated summaries. As a result, these approaches produce short (3-6 sentences), inflexible, and flat summaries that are the same for all users. Consequently, these approaches fail to provide more extensive summaries even when users express interest in obtaining additional information.

\textbf{Reference Summaries.} Traditional document summarization techniques rely on reference summaries created by humans for training their systems. However, this approach is subjective and, more importantly, resource-intensive. For instance, Lin \cite{lin2004rouge} reported that creating summaries for the Document Understanding Conferences (DUC) required 3,000 hours of human effort. Personalized summaries eliminate the need for such reference summaries by generating specific summary for a user instead of optimizing a summary for all users.

\textbf{Our Contribution.} 
%  to efficiently convey the key ideas of documents, 
We study the automatic creation of personalized, structured summaries, allowing the user to overview a document collection's content without much reading quickly. 
The goal here is to dynamically maintain a federated summary view incrementally, resulting in a unified framework for intelligent summary generation and data discovery tools from a user-centered perspective. 
The unique contribution of this paper includes:
\begin{itemize}
    \item We provide summaries in the form of a \textit{hierarchical concept map}, labeled graphs representing concepts and relationships in a visual and concise format.
    Their structured nature can reveal interesting patterns in documents that users would otherwise need to discover manually.
It enables providing more information than traditional approaches within the same limit size. 
    It can be used as a navigator in the document collection.
    Such visualization is beneficial for decision-making systems.
    \item We introduce and formalize a theoretically grounded method.
    We propose a personalized interactive summarization approach utilizing a reinforcement learning algorithm to learn generating user-adapted results.
    It is the first approach to predict users' desired structured summary to the best of our knowledge.
    % It also has an additional advantage, which is being interpretable.
    % Therefore, 
    \item We provide various evidence evaluating different aspects to prove \textit{Summation}'s usability using human and automatic evaluation.
\end{itemize}
\begin{figure*}[t]    \centerline{\includegraphics[width=0.9\textwidth]{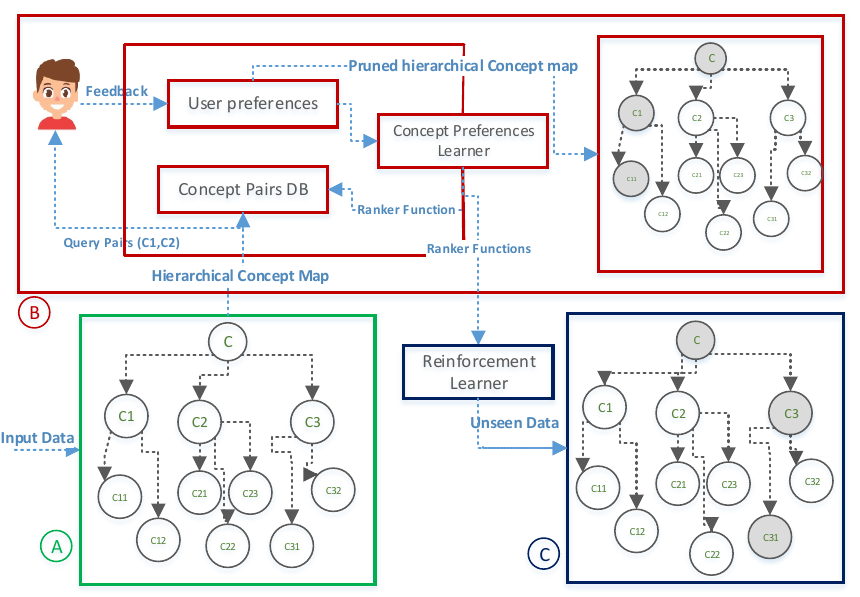}}
    \caption[]{An overview of \textit{Summation}: A) The input data is converted to a hierarchical concept map (organizer). B) The summarizer is responsible for extracting and learning user preferences. C) Using reinforcement learning to predict users' desired summary.}
    \label{overview}
\end{figure*}
We divide the proposed framework into two steps.
The first step is \textit{organizer} which structure unstructured data by making a hierarchical concept map.
Then \textit{summarizer} is responsible for: i) predicting users' preferences based on the given feedback by employing \textit{preference learning} and ii) learning to provide personalized summaries by leveraging reinforcement learning.
A general overview of the algorithm is depicted in Figure~\ref{overview}.

\section{Related Work}
\label{Related}
We categorize previous approaches into three groups including \textit{traditional approaches}, \textit{structured approaches}, \textit{personalized and interactive approaches} discussed below.

\textbf{Traditional Approaches.}
A good summary should provide the maximum information about the input documents within a size limit and be fluent and natural.
Different aspects for categorizing traditional multi-document summarization approaches exist, such as the input type, the process, and the summarization goal~\cite{gupta2010survey,ghodratnama2021rare}.
However, the main category considers the process and the output type of the summarization algorithm: \textit{extractive} and \textit{abstractive} approaches. 
The input in both cases is a set of documents, and the output is a few sentences.
Abstractive summaries are generated by interpreting the main concepts of a document and then stating those contents in another format.
Therefore, abstractive approaches require deep natural language processing, such as semantic representation and inference~\cite{gupta2010survey}.
However, extractive text summarization selects some sentences from the original documents as the summary.
These sentences are then concatenated into a shorter text to produce a meaningful and coherent summary~\cite{mehta2018effective}.
Early extractive approaches focused on shallow features, employing graph structure, or extracting the semantically related words~\cite{edmundson1969new}. 
Different machine learning approaches, such as naive-Bayes, decision trees, neural networks, and deep reinforcement learning models are used for this purpose~\cite{ghodratnama2020extractive,wu2018learning,narayan2018ranking}.

\textbf{Structured Approaches.}
While traditional summarization approaches produce unstructured summaries, there exist few attempts on structured summaries.
Structured summaries are defined by generating Wikipedia articles and biographies to extract the significant aspects of a topic using approaches such as topic modeling or an entity-aspect LDA model~\cite{liu2010biosnowball,ghodratnama2015efficient}.
Discovering threads of related documents is another category of structured summaries.
They mostly use a machine algorithm to find the threads using a supervised approach and features such as temporal locality of stories for event recognition and time-ordering to capture dependencies \cite{nallapati2004event}. 
A few papers have examined the relationship between summarization and hierarchies.
However, the concept of hierarchy in these approaches is the relation between different elements of a document. 
An example is creating a hierarchy of words or phrases to organize a set of documents~\cite{haghighi2009exploring}.
There is a related thread of research on identifying the hierarchical structure of the input documents and generating a summary which prioritizes the more general information according to the hierarchical structure~\cite{christensen2014hierarchical}.
% celikyilmaz2010hybrid,
However, the information unit is a sentence, and the hierarchy is based on time measures.
Concept-based multi-document summarization is a variant of traditional summarization that produces structured summaries using concept maps.
It learns to identify and merge coreferent concepts to reduce redundancy and finds an optimal summary via integer linear programming.
However, it produces a single flat summary for all users~\cite{falke2019automatic}.

\textbf{Personalized and Interactive Approaches.}
Recently, there exist few recent attempts on personalized and interactive approaches in different NLP tasks.
Unlike non-interactive systems that only present the system output to the end-user, interactive NLP algorithms ask the user to provide certain feedback forms to refine the model and generate higher-quality outcomes tailored to the user.
Multiple forms of feedback also have been studied including mouse-clicks for information retrieval~\cite{borisov2018click}, post-edits and ratings for machine translation~\cite{borisov2018click,kreutzer2018can}, error markings for semantic parsing~\cite{lawrence2018counterfactual}, and preferences for translation~\cite{kingma2014adam}. 
A significant category of interactive approaches presents the output of a given automatic summarization system to users as a draft summary, asking them to refine the results without further interaction.
The refining process includes cutting, paste, and reorganize the essential elements to formulate a final summary~\cite{orasan2006computer,narita2002web}.
% craven2000abstracts
% Other works present automatically derived hierarchically ordered summaries allowing users to drill down from a general overview to detailed information~\cite{christensen2014hierarchical,shapira2017interactive}.
% Therefore, these systems are neither interactive nor consider the user's feedback to update their internal summarization models.
Other interactive summarization systems include the iNeATS~\cite{leuski2003ineats} and IDS~\cite{jones2002interactive} systems that allow users to tune several parameters for customizing the produced summaries.
Avinesh and Meyer~\cite{avinesh2017joint} proposed the most recent interactive summarization approach that asks users to label important bigrams within candidate summaries.
Their system can achieve near-optimal performance.
However, labeling important bigrams is an enormous burden on the users, as users have to read through many potentially unimportant bigrams.
Besides, it produces extractive summaries that are unstructured.

\section{The proposed Approach (Summation)}
\label{proposed}
The ultimate goal of summarization is to provide a concise, understandable, and interpretable summary tailored to the users' needs.
However, making such a summary is challenging due to massive document collection, the speed of generated documents, and the unstructured format.
% make it hard to extract the most relevant information and omit redundancy satisfying a length constraint.
In this regard, \textit{Summation} aims to make structured summaries to facilitate further processes to make it concise and easily understandable while engaging users to create their personalized summaries.
This novel framework has two components: \textit{organizer} and the \textit{summarizer}.
First, we discuss the problem definition, and then each component is explained.

\textbf{Problem Definition.}
The input is a set of documents $D=\{D_{1},D_{2}, ... ,D_{N}\}$ and each document consists of a sequence of sentences $S=[s_1,s_2,$$...$$,s_n]$. 
Each sentence $s_i$ is a set of concepts $\{c_1,c_2, ..,c_k\}$, where a concept can be a word (unigram) or a sequence of words.
The output is a personalized hierarchical concept map.
This novel framework has two components, an organizer and a summarizer, explained in Sec.~\ref{sum_Organizer} and \ref{sum_summarizier}, respectively.
% Input to the \textit{Summation} can be either a single document or a document cluster on the same topic.
% Input $X$ is a set of related documents, and $x$ is a cluster of documents.
% $Y(x)$ is all summaries can be built in the form of a concept map for the document cluster $x$, and $y \in Y(x)$ is a potential summary with limited size $b$ defined by the user.
% %where concepts are con
% The goal of \textit{Summation} is to map each input document cluster $x$ to its best summary in $Y(x)$ for a specific user $u$.
% We formulate the problem as a sequential decision-making problem modeled as an episodic Markov Decision Process (MDP).
% Concepts are selected based on the learned user preference by interacting with users and are added to the draft summary to form the user's desired summary.
% \vspace{-2mm}
\subsection{Adding Structure to Unstructured Data}
\label{sum_Organizer}
The first step is to structure unstructured information by making a hierarchical concept map.
A concept map is a graph with directed edges, where nodes indicate concepts and edges indicate relations.
Both concepts and relations are sequences of related words representing a semantic unit.
Consequently, the first step in creating a concept map is to identify all concepts and relations.
Here, we propose hierarchical clustering to form the hierarchical concept map.
% Abstract labels are created to make summaries concise and coherent.
% , and abstract labels are created to make summaries concise and coherent.
% \vspace{-4mm}
\subsubsection{Concept and Relation Extraction.}
Concepts come in different syntactic types, including nouns, proper nouns, more complex noun phrases, and verb phrases that describe activities~\cite{falke2019automatic}.
For this purpose, we used open information extraction (OIE)~\cite{etzioni2008open} through which the entities and relations are obtained directly from the text.
OIE finds binary propositions from a set of documents in the form of ($con_1$,$R$,$con_2$), which are equivalent to the desired concepts and relations. 
For example, the output for the sentence, ‘cancer treatment is underpinned by the Pharmaceutical Benefits Scheme’, is:
%show it as image
\textit{Cancer treatment$\xrightarrow[\text{}]{\text{is underpinned}}$ by the Pharmaceutical Benefits Scheme}

% \begin{figure*}[h]
%     \centerline{\includegraphics[width=0.8\textwidth]{exampleconcept.pdf}}
%     \caption[]{An overview of the proposed approach (Adaptive Summaries). 1) Summaries are initiated with ExDos~\cite{samira2020}. 2) Users integrate their preferences in making summaries by giving feedback in an iterative loop. 3) An example of user interaction.}
%     \label{overview}
% \end{figure*}
% Different challenges in extracting concepts and relations exist.
Balancing precision and recall in extracting concepts is a challenging task.
A high precision causes to define all identified spans as mentions of concepts.
Therefore, some constructions are usually missed, which leads to lowering the recall.
On the other hand, a high recall is necessary since missed concepts can never be in summary.
Obtaining a higher recall may extract too many mentions, including false positives.
% Besides, an extensive set of extracted mentions makes selecting summary-worthy elements harder in terms of the number of options. 
% Generalizability is also essential as extracting spans of a specific syntactic structure might yield only correct mentions on a particular text.
Generalizability is also essential.
The reason is that extracting a particular syntactic structure might generate only correct mentions, causing too broad mentions.
Ideally, a proper method applies to many text types. 
To avoid meaningless and long concepts, we processed the OIE results such that concepts with less than one noun token or more than five tokens are omitted.
The original nouns also replace pronouns.
If an argument is a conjunction indicating conj-dependency in the parse tree, we split them.
%on the text it was designed for
% However, it may be too broad, covering many undesired spans on other text types.
% Ideally, a proper method applies to many text types.
% To avoid meaningless and long concepts, we process OIE's results such that each concept should contain at least one noun token and not be longer than five tokens.
% Pronouns are laso replaced by their original nouns.
% Besides, if an argument is a conjunction indicating conj-dependency in the parse tree, we split them.
% If the second argument start with verb we change the position to to relation.
% In addition, concepts are often referred to using pronouns, as in sentence (6) of the example.
% \vspace{-2mm}
\subsubsection{Concept Map Construction.}
Among various extracted concepts and relations, multiple expressions can refer to the same concept while not using precisely the same words; that is, they can also use synonyms or paraphrases.
However, distinguishing similar concepts to group them is challenging and subjective.
For example, adding a modifier can completely change the meaning of a concept based on the purpose of summarization.
Consequently, grouping them may lead to propositions that are not stated in the document.
Therefore, we need to group every subset that contains mentions of a single, unique concept.
Scalability is another critical issue.
For example, pairwise comparisons of concepts cause a quadratic run-time complexity applicable only to limited-sized document sets.
The same challenges exist for relation grouping.
However, we first grouped all mentions by the concepts' pairs, and then performed relation grouping.
Therefore, this task’s scope and relevance are much smaller than when concepts are used.
Therefore, in practise, comparison-based quadratic approaches are feasible.
Moreover, as the final goal is to create a defined size summary, the summary size significantly affects the level of details in grouping concepts. 
This is because the distinction between different mentions of a concept might not be required, as it is a subjective task.
Ideally, the decision to merge must be made based on the final summary map’s propositions to define the necessary concept granularity.

We further propose hierarchical conceptual clustering using k-means with word embedding vectors to tackle this problem, as it spans a semantic space. 
Therefore, word embedding clusters give a higher semantic space, grouping semantically similar word classes under the Euclidean metric constraint defined below.
Before defining the proposed hierarchical conceptual clustering, we review word embedding schemes used in the proposed model. 
% We propose a hierarchical conceptual clustering using k-means with word embedding vectors to tackle this problem as it spans a semantic space.
% Therefore, word embedding clusters gives a higher semantic space, grouping semantic similar word classes under the Euclidean metric constraint defined below.
% Before defining the proposed hierarchical conceptual clustering, we review word embedding schemes used in this paper.

\textbf{Word Embedding.}
Word embedding is a learnt representation of text such that the same meaning words have similar representations.
Different techniques can be used to learn a word embedding from the text.
Word2Vec~\cite{mikolov2013distributed} is an example of a statistical model for learning a word embedding representation from a text corpus, utilising different architectures.
As such, we used skip-gram and bag of character n-grams in our experiments.
The skip-gram model uses the current word for predicting the surrounding words by increasing the weights of nearby context words more than other words using a neural network model.
One drawback of skip-gram is its inability to detect rare words.
In another model, authors define an embedding method by representing each word as the sum of the vector representations of its character n-grams, known as ‘bag of character n-grams’~\cite{bojanowski2017enriching}.
If the training corpus is small, character n-grams will outperform the skip-gram (of words) approach.~\footnote{We used fastText for word embedding:  https://fasttext.cc/docs/en/support.html}

\begin{figure*}[t]
    \centerline{\includegraphics[width=0.7\textwidth]{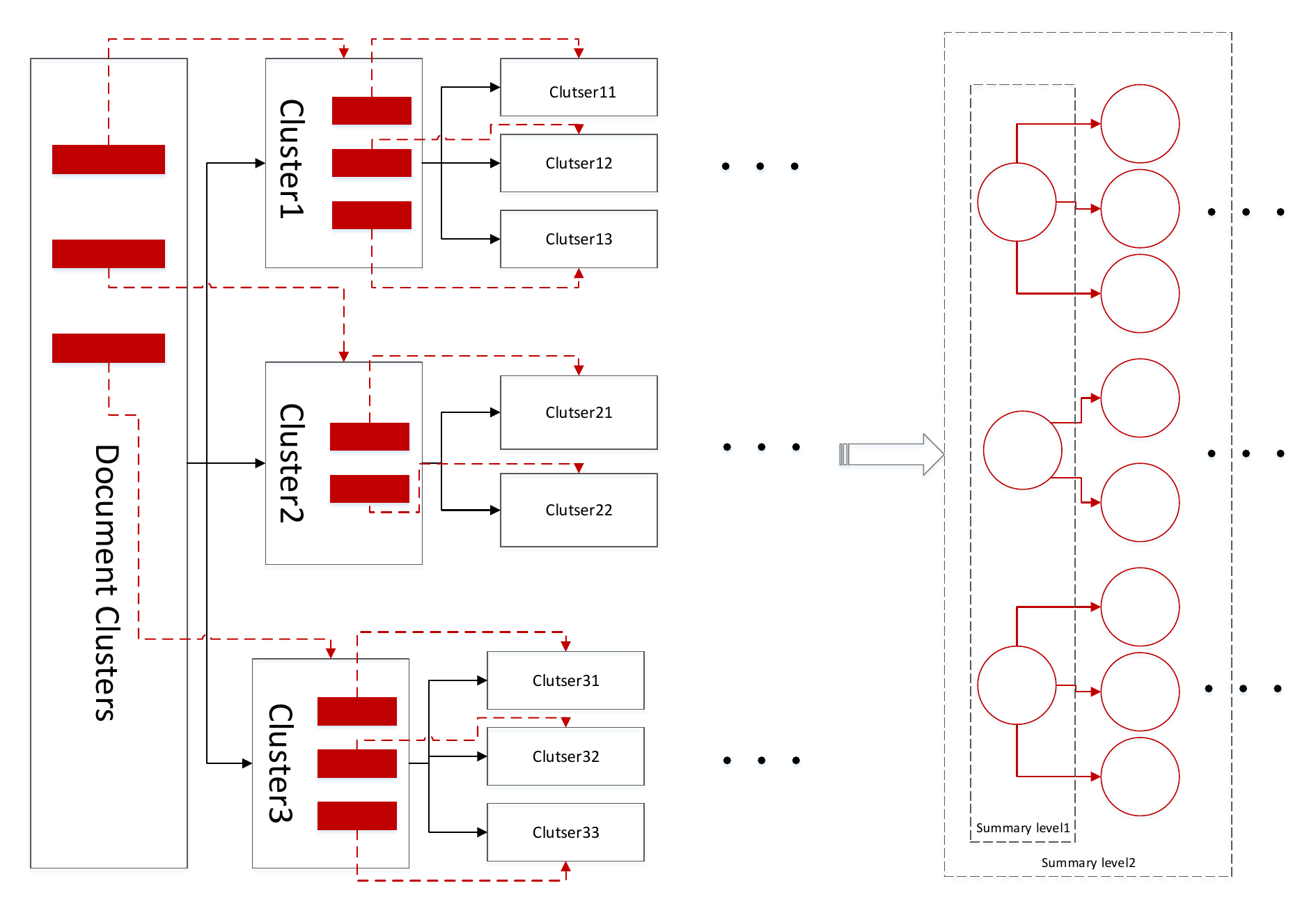}}
    \caption[]{The hierarchical conceptual clustering architecture.}
    \label{heirarchical}
\end{figure*}

\textbf{Conceptual Hierarchical Clustering.} Given word (concept) embeddings learnt from a corpus, $\{v_{w_1},v_{w_2},...,v_{w_T}\}$, we propose a novel recursive clustering algorithm to form a hierarchical concept map, $H$.
This variable denotes a set of concept maps organised into a hierarchy that incrementally maintains hierarchical summaries from the most general node (root) to the most specific summary (leaves).
Within this structure, any non-leaf summary generalises the content of its children nodes. 
Hierarchical summarization has two critical strengths in the context of large-scale summarization.
First, the initial information under review is small and grows upon users’ request, so as not to overwhelm them.
Second, the parent-to-child links facilitate user navigation and drilling down for more details on interesting topics.
The hierarchical conceptual clustering minimizes the objective function Eq.~\ref{clustering2} over all k clusters as C=\{$c_1,c_2,..,c_k$\}. 
% and $k=b$ using local search algorithm. 
\begin{equation}
\label{clustering2}
J = \sum_{k=1}^{K}\sum_{t=1}^{\mid T \mid} |v_{w_t}- c_k|^2 +\alpha \min _{c\in C} size(c),
\end{equation}
where $c_k$  is the randomly selected centre $k-th$ cluster, and $T$ is the number of word vectors.
The second term is the evenness of the clusters, added to avoid clusters with small sizes.
$\alpha$ tunes the evenness factor, which was defined by employing a grid search over a development set.
We also implemented hierarchical clustering top-down at each time, optimising Eq.~\ref{clustering2}.
After defining the clusters, we must find the concept that best represents every concept at the lower levels to ensure hierarchical abstraction. 
A concise label is the desired label for each node; however, shortening mentions can introduce propositions that are not asserted by a text.
For example, the concept labelled ‘students’ can change in meaning where the emphasis is on a few students or some students.
To this end, a centre of a cluster at each level of the hierarchy was defined as a label.
The inverse distance to the cluster centres is the membership degree or the similarity to each label.
The cluster distance for a word $w_t$ is defined as $d_{v_{w_t}}$.
Consequently, the membership of each word $w_t$ in cluster $c_k$ to its label is the inverse distance defined in Eq.~\ref{ch6_eq20}.
\begin{equation}
\label{ch6_eq20}
m_{v_{w_t}}=\frac{1}{d_{v_{w_t}}} =\frac{1}{|c_k-v_{w_t}|^2} \hspace{2mm} \forall w_t \in c_k
\end{equation}

We then fine-tuned K within the 5–50 range based on the dataset size and chose the cluster number according to gap statistic value~\cite{tibshirani2001estimating}.
The output $H$ can be directly used as a new dataset for other actions, such as browsing, querying, data mining process, or any other procedures requiring a reduced but structured version of data.
The hierarchical clustering can also be pruned at each level to represent a summarised concept map for different purposes or users.
Therefore, $H$ is fed to the summariser for pruning to generate a personalized summary. Moreover, by using preference-based learning and RL, we learn users’ preferences in making personalized summaries for unseen topic-related documents, discussed in Sec.~\ref{sum_summarizier}.

\subsection{Summarizer}
\label{sum_summarizier}
The hierarchical concept map produced in the previous step is given to the summariser to make the desired summaries for users based on their given preferences.
Therefore, the summariser consists of two phases\textemdash(i) predicting user preferences and (ii) generating the desired summary.
% \vspace{-2mm}

\subsubsection{Predicting User Preference.}
% The first step towards creating Personalized summaries is to understand users’ interests.
% The same procedure in SumRecom is used for extracting users’ preferences; however, the selection of sentences is among hierarchy nodes.
The first step towards creating personalized summaries is to understand users’ interests.
It can be extracted implicitly based on users' profiles, browsing history, likes or dislikes, or retweeting in social media~\cite{alhindi2015profile}.
When this information is not available, interaction with users is an alternative to retrieve user's perspectives.
The user feedback can be in any form, such as mouse-click or post-edits, as explained in Section~\ref{Related}.
Preference-based interactive approaches are another form of feedback that puts a lower cognitive burden on human subjects~\cite{zopf2018estimating}.
% ,kingsley2010preference
% Concept selection aims to find the desired information within a given set of source documents.
% s than asking for absolute ratings or categorized labels as it is a binary decision
For instance, asking users to select one concept among ``cancer treatment" and ``cancer symptoms" is more straightforward than asking for giving a score to each of these concepts.
Therefore, in this paper, to reduce users' cognitive load, queries are in the form of concept preference.
% Since humans can quickly assess the importance of concepts given a topic, we query users to ask preference over concepts. 
% Active learning is also used to reduce the number of interaction rounds.
% However, in 
% To recap, we use active preference learning (APL) in an interaction loop to maximize the information gained from a small number of preferences, reducing the sample complexity.
% In the following, the active and preference learning implemented in the proposed method is discussed.
Preference learning is a classification method that learns to rank instances based on the observed preference information.
It trains based on a set of pairwise preferred items and obtaining the total ranking of objects~\cite{furnkranz2010preference}.
% Typical usage is in ranking search results according to the feedback of user preference. 

$H$ is the hierarchical concept map, where at the $i-th$ level of the hierarchy there exist $m_i$ nodes defining a label $l$.
$L=\{l_{11},...,l_{nm_i}\}$ is the set of all labels, where $l_{i1}$ indicates the first node at $i-th$ level of the hierarchy and $n$ is the number of levels, and $L_i$ indicates the labels at  $i-th$ level.
We queried users with a set of pairwise concepts at the same levels,
$\{p(l_{i1},l_{i2}),p(l_{i2},l_{i3}),...,p(l_{im_i-1},p(l_{im_i})\}$, where $p(l_{i1},l_{i2})$ is defined in Eq.~\ref{ch6_eq21}.
% $H$ is the hierarchical concept map, where at $i-th$ level of the hierarchy, there exist $m_i$ nodes defining a label $l$.
% $L=\{l_{11},...,l_{nm_i}\}$ is the set of all labels where $l_{i1}$ indicates the first node at $i-th$ level of the hierarchy and $n$ is the number of levels and $L_i$ indicates the labels at $i-th$ level.
% We query users a set of pairwise concepts at same levels, $\{p(l_{i1},l_{i2}),p(l_{i2},l_{i3}),...,p(l_{im_i-1},p(l_{im_i})\}$, where $p(l_{i1},l_{i2})$ is defined as:
% \vspace{-1mm}
\begin{equation}
\label{ch6_eq21}
  p(l_{i1},l_{i2})=\begin{cases}
    1, & \text{if $l_{i1}>l_{i2}$}\\
    0, & \text{otherwise}
  \end{cases}
\end{equation}
where $>$ indicates the preference of $l_{i1}$ over $l_{i2}$.
Preference learning aims to predict the overall ranking of concepts, which requires transforms concepts into real numbers, called utility function.
The utility function $U$ such that $ l_i > l_j \xrightarrow{} U(l_i) > U (l_j)$, where $U$ is a function $U: C\xrightarrow{} \mathbbm{R}$.
In this problem, the ground-truth utility function ($U$) measures each concept’s importance based on users’ attitudes, defined as a regression learning problem.
According to $U$, we defined the ranking function, $R$, measuring the importance of each concept towards other concepts based on users’ attitude.
This is defined in Eq.~\ref{ch6_eq22}.
\begin{equation}
\label{ch6_eq22}
    R(l_{i})=\sum \mathbbm{1} \{U(l_{i})>U(l_{j})\} , \forall l_{i} , l_{j} \in {L}
\end{equation}
where $\mathbbm{1}$ is the indicator function.
The Bradley–Terry model~\cite{szummer2011semi,zakershahrak2020we} is a probability model widely used in preference learning.
Given a pair of individuals $l_{i}$ and $l_{j}$ drawn from some population, the model estimates the probability that the pairwise comparison $l_{i} > l_{j}$ is true.
Having $n$  observed preference items, the model approximates the ranking function $R$ by computing the maximum likelihood estimate in Eq.~\ref{ch6_eq23}.
\begin{equation}
\label{ch6_eq23}
\begin{split}
J_x(w)= & \sum_{i \in n}[p(l_{i},l_{j})log F(l_{i},l_{j};w)+
p(l_{j},l_{i})log F(l_{j},l_{i};w)]
\end{split}
\end{equation}
where $F(l)$ is the logistic function defined in Eq.~\ref{ch6_eq24}.
\begin{equation} \label{ch6_eq24}
 F(l_{i},l_{j};w)= \frac{1}{1+exp[U^{*}{(l_j;w)}-U^{*}{(l_i;w)}]}
\end{equation}

Here, $U^{*}$ is the approximation of $U$ parameterised by $w$, which can be learnt using different function approximation techniques.
In our problem, a linear regression model was designed for this purpose, defined as $U(l;w)=w^{T}\phi(l)$, where $\phi(l)$ is the representation feature vector of the concept $l$.
For any $l_i,l_j \in L$, the ranker prefers $l_i$ over $l_j$ if $w^{T}\phi(l_i)> w^{T}\phi(l_j)$.

By maximizing the $J_x(w)$ in Eq.~\ref{ch6_eq23}, $w^{*} = arg max_w J_x(w)$, the resulting $w^{*}$ using stochastic gradient ascent optimisation will be used to estimate $U^{*}$, and consequently the approximated ranking function $R^{*}: C \xrightarrow[]{} \mathbbm{R}$.
Thus, \textit{Summation} learns a ranking over concepts and uses the ranking to generate personalized summaries.

% \vspace{-2mm}
\subsubsection{Generating Personalized Summaries.}
The summarization task is to transform the input (a cluster of documents) $d$ to the best summary among all possible summaries, called $Y(d)$, for the learnt preference ranking function.
This problem can be defined as a sequential decision-making problem, starting from the root, sequentially selecting concepts and adding them to a draft summary.Therefore, it can be defined as an MDP problem.

An MDP is a tuple $(S,A,R,T)$, where $S$ is the set of states, $A$ is the set of actions, $R(s,a)$  is the reward for performing an action ($a$) in a state ($s$), and $T$ is the set of terminal states.
In our problem, a state is a draft summary, and $A$ includes two types of action\textemdash either adding a new concept to the current draft summary or terminating the construction process if it reaches users’ limit size.
The reward function $R$ returns an evaluation score in one of the termination states or $0$ in other states.

A policy $\pi(s,a): S \times A \xrightarrow{} R$ in an MDP defines the selection of actions in state $s$.
The goal of RL algorithms is to learn a policy that maximises the accumulated reward.
The learnt policy trained on specific users’ interests is used on unseen data at the test time (in this problem to generate summaries in new and related topic documents).

We defined the reward as the summation of all concepts’ importance included in the summary.
A policy$\pi$  defines the strategy to add concepts to the draft summary to build a user’s desired summary. 
We defined $\pi$ as the probability of choosing a summary of $y$ among all possible summaries within the limit size using different hierarchy paths, $Y(d)$, denoted as $\pi(y)$. 
The expected reward of performing policy $\pi$, where $ R(y)$ is the reward for selecting summary $y$, is defined in Eq.~\ref{ch6_eq25}.

\begin{equation}
\label{ch6_eq25}
    R^{RL}(\pi|d)= \mathbbm{E}_{y \in Y(d)}R(y)= \sum_{y\in Y(d)} \pi(y)R(y)
    % =\sum_{y\in Y} \pi(y)R(y)
\end{equation}

\begin{algorithm}[t]
\caption{Summation}
\label{alg1}
\begin{algorithmic}[1]
\Statex \textbf{Input:} Document cluster $d$
\Statex \textbf{Output:} Summary ($H$) and optimal policy

\Procedure{Summation}{}
\State \textbf{Organiser}:
\State $Concepts\ and\ Relations \gets$ Concept and relations extraction ($d$)
\State $H \gets$ Hierarchical conceptual clustering ($\textit{Concepts and Relations}$)

\State \textbf{Summarizer (User preference learner (iteratively))}:
\State $User\ preferences \gets$ Query pairs (user)
\State $Ranker\ function \gets$ Preference learner ($\textit{User preferences}$)

\State \textbf{Summarizer (RL learner)}:
\State $Optimal\ policy \gets$ Policy learner ($\textit{Ranker function}$)

\State \Return Summary ($H$) and optimal policy
\EndProcedure

\end{algorithmic}
\end{algorithm}

% \begin{algorithm}[t]
% \caption{Summation}
% \begin{flushleft}
% \textbf{Input}: Document Cluster x. \\
% \textbf{Output}: Summary(H) and Optimal Policy. 
% \end{flushleft}
% \Procedure{Summation.}{}
% \begin{algorithmic}
% \label{alg1}
% \STATE\emph{Organizer}:
%     \State $\textit{Concepts and Relations} \gets \textit{Concept and Relation Extraction (x)}$
%      \State $\textit {H} \gets \textit{Hierarchical Conceptual Clustering (Concepts)}$
% \STATE\emph{Summarizer (User Preference Learner)}:
%     \State $\textit{User Preferences} \gets \textit{Query pairs (user)}$
%     \State $\textit{Ranker Function} \gets \textit{Preference Learner (User Preferences)}$
% \STATE\emph{Summarizer (RL Learner)}:
%     \State $\textit{Optimal Policy} \gets \textit{Policy Learner ( Ranker Function)}$
% \RETURN Summary(H) and Optimal Policy
% \end{algorithmic}
% \end{algorithm}
% \vspace{-2mm}
The goal of MDP is to find the optimal policy $\pi^*$ that has the highest expected reward.
Therefore, the optimal policy, $\pi^*$, is the function that finds the desired summary for a given input based on user feedback (Eq.~\ref{ch6_eq26}).

\begin{equation}
\label{ch6_eq26}
\pi^* = arg max \hspace{2mm} R^{RL}(\pi|d) = arg max \sum_{y \in Y(d)}\pi(y) R(y)
\end{equation}
We also used the linear temporal difference algorithm to obtain $\pi^*$.
The process is explained in Algorithm~\ref{alg1}.

\section{Evaluation}
\label{Evaluation}
In this section, we present the experimental setup for assessing our summarization model's performance.
We discuss the datasets, give implementation details, and explain how system output was evaluated.
% \vspace{-5mm}
\subsection{Datasets and Evaluation}
We evaluated \textit{Summation} using three commonly employed benchmark datasets from the Document Understanding Conferences (DUC) \footnote{Produced by the National Institute Standards and Technology (https://duc.nist.gov/)}.
% and CNN/Daily Mail~\footnote{ https://github.com/abisee/cnn-dailymail}.
% \cite{hermann2015teaching}
% The latter contains news articles (781 tokens on average) paired with multi-sentence summaries (3.75 sentences or 56 tokens on average).
% , 287,226 training pairs, 13,368 validation, and 11,490 test pairs.
% Details are described in Table~\ref{tab:dataset}.
Each dataset contains a set of document clusters accompanied by several human-generated summaries used for training and evaluation. 
Details are explained in Table~\ref{tab:dataset}
% The document numbers in DUC1, DUC2, and DUC4 are respectively 30, 58, and 50.
% The document clusters are 308, 567, and 500, and the average number of sentences in each document is 378, 271, and 265.
% We randomly split both datasets into equally sized training and test sets.
% We  evaluate  our  approach  using  two  benchmarkdatasets and compare the generated concept mapsagainst  reference  maps.   As  the  first  dataset,  weuse  a  recently  published  corpus  by  Falke  andGurevych (2017a) that provides summary conceptmaps for document clusters on educational topics.They were manually created using crowdsourcingand expert annotators.  As the second dataset, weuse a corpus in which the introductions of featuredWikipedia articles are used as summaries for webdocuments (Zopf et al., 2016).  This property al-lows us to make use of the links to other Wikipediapages in the summaries as annotations of concepts.In combination with Open Information Extraction,we  can  therefore  automatically  derive  conceptsand relations from the Wikipedia summaries to ob-tain a second corpus of summary concept maps.We refer to these datasets asEDUCandWIKI.Table 1 shows their characteristics.   Note that inboth datasets the summaries are much smaller thanthe document sets, posing a challenging summa-rization  task.In  addition,  the  document  clus-ters of  EDUCare very large,  constituting a chal-lenging but real-world evaluation setting regardingcomputational efficiency.  We randomly split bothdatasets into equally sized training and test sets
% \subsection{Evaluation Metric}
\begin{table}[t]
\centering
  \centering \caption{Dataset description: indicating the number of documents, number of document clusters, and the average number of sentences in each document.}
  % \vspace{1mm}
  \label{tab:dataset}
  \tabcolsep=0.11cm
  \begin{tabular}{|c|c|c|l|}
    \hline
    Dataset & Doc-Num & Cluster-Num & Sentence\\
  \hline
    DUC1 &30&308&378\\
    \hline
    DUC2 &59&567&271\\
    \hline
    DUC4 &50&500&265\\
    \hline
    % CNN/Daily &287,226(train) &&39.87(781 tokens) \\
    %  \hline
\end{tabular}
\end{table}

\textbf{Automatic Evaluation.} We evaluate the quality of summaries using $ROUGE_N$ measure~\cite{lin2004rouge}\footnote{We run ROUGE 1.5.5: http://www.berouge.com/Pages/defailt.aspx with parameters -n 2 -m -u -c 95 -r 1000 -f A -p 0.5 -t 0} defined as:

The three variants of ROUGE (ROUGE-1, ROUGE-2, and ROUGE-L) are used.
% ROUGE-1 and ROUGE-2 are used to evaluate informativeness, and ROUGE-L (longest common subsequence) is used to evaluate the fluency.
We used the limited length ROUGE recall-only evaluation (75 words) to avoid being biased. 

\textbf{Human Evaluation.} For this purpose, we hired fifteen Amazon Mechanical Turk (AMT)\footnote{https://www.mturk.com/} workers to attend tasks without any specific prior background required.
Then five document clusters are randomly selected from the DUC datasets. 
%We design a series of micro-tasks for each experiment.
%We selected not recently published articles to avoid any bias in understanding topics.
Each evaluator was presented with three documents to avoid any subjects' bias and was given two minutes to read each article.
To make sure human subjects understood the study's objective, we asked workers to complete a qualification task first.
They were required to write a summary of their understanding.
We manually removed spam from our results.
% Besides, the full-length F1 score is used for the evaluation of the CNN/DailyMail dataset.
%As the experiment, the target is to create a summary with at most $100$ words for each cluster.
% To decide the best parameters, we perform10-fold cross validation on DUC’01.
% \vspace{-2mm}
\subsection{Results and Analysis}
% First, we explain the evaluation settings, and then we discuss the results and analysis.
\textit{Summation} was evaluated from different evaluation aspects, first from the organiser’s output, and then concerning the hierarchical concept map ($H$), which can be served individually to users as the structured summarised data.
Next, we evaluated $H$ using both human and automatic evaluation techniques to answer the following questions:

\begin{itemize}
    \item Do users prefer hierarchical concept maps to explore new and complex topics?
    \item How much do users learn from a hierarchical concept map?
    \item How coherent is the produced hierarchical concept map?
    \item How informative are summaries in the form of a hierarchical concept map?
\end{itemize}
Personalized summaries generated on test data were also evaluated from various perspectives to analyse the effect of RL and preference learning, including:
\begin{itemize}
    \item The impact of different features in approximating the proposed preference learning.
    \item The role of the query budget in retrieving pairwise preferences.
    \item The performance of RL algorithm and the information coverage in terms of ROUGE.
    \item Users' perspectives on learned summaries based on their given feedback.
    % \item The effect of clustering in removing redundancy.
    % \item The effect of different values for number of clusters 
\end{itemize}

\textbf{Hierarchical Concept Map Evaluation.}
To answer the questions in Sec.~\ref{Evaluation}, we performed three experiments. 
First, within the same limit size as the reference summaries, we compared the summaries produced by three models\textemdash using ExDos, which is a traditional approach; using a traditional hierarchical approach~\cite{christensen2014hierarchical}; and using a structured summarization approach~\cite{falke2019automatic} on selected documents (with ROUGE-1 and ROUGE-2 scores based on the reference summaries).
The average ROUGE-1 for Summation was 0.65 and ROUGE-2 was 0.48.
The structured approach~\cite{falke2019automatic} showed similar performance with ROUGE-1 and ROUGE-2 at 0.65 and 0.45, respectively.
Meanwhile, traditional hierarchical approaches~\cite{christensen2014hierarchical} produced a ROUGE-1 of 0.27 and ROUGE-2 of 0.18.
In the same task, the percentage of covered unigrams and bigrams based on documents were also compared.
Both Summation and the structured approach covered approximately 4\% unigrams and 2\% bigrams, but dropped below 1\% in both cases when testing the hierarchical approaches.
In the third experiment, all competitors’ outputs were rated based on three measures, including usability in exploring new topics, level of informativeness, and coherency.
Summation’s rate for the first and second criteria was 96\% and 94\%, respectively.
However, it was 34\% for coherency.
We removed all concepts with low similarity to their parents based on a different threshold at each level.
After repeating the same experiment, and rate of coherency increased to 76\%.

\begin{figure}[t]
\centering
    \includegraphics[width=\linewidth]{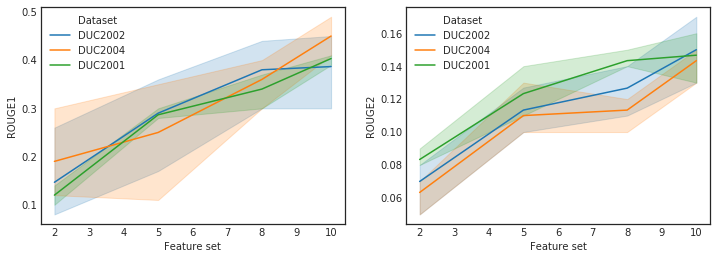}
    \caption{Evaluating different feature set for estimating the ranker function.}
    \label{featureAnalysis}
\end{figure}

\textbf{Feature Analysis.}
\label{feature}
Before evaluating the effect of conceptual preference, it is important to explain the ground-truth concept ranker function ($U$) and the approximate function ($U^*$), indicating the importance of concepts.
To estimate the approximate function ($U^*$), we defined a linear model $U^*(c)=W^T\phi(c)$, where $\phi$ are the features.
To this end, a set of features (whose importance was validated in ExDos) was used, including surface-level and linguistic-level features.
Surface-level features include frequency-based features (TF-IDF, RIDF, gain and word co-occurrence), word-based features (upper-case words and signature words), similarity-based features (Word2Vec and Jaccard measure) and named entities.
Linguistic features are generated using semantic graphs and include the average weights of connected edges, the merge status of concepts as a binary feature, the number of concepts merged with a concept, and the number of concepts connected to the concept.
We defined different combinations of features with different sizes,$\{2,5,8,10\}$, starting from the most critical one.
Then, we repeated the experiments for 10 cluster documents.
We used the concepts included in the reference summary as preferences, and then evaluated the concept coverage in a concept map compared to the reference summaries using ROUGE-1 and ROUGE-2.
The results reported in Fig.~\ref{featureAnalysis} show that the model’s performance improved after adding more features.

\begin{figure}[t]
\centering
    \includegraphics[width=\linewidth]{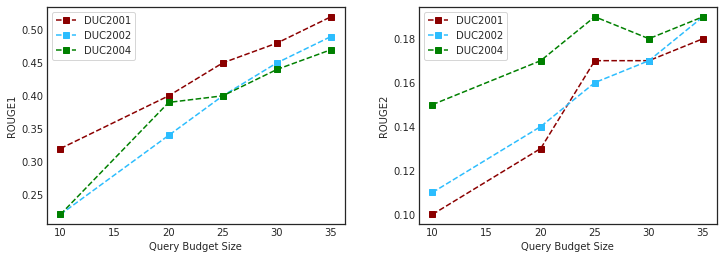}
    \caption{The effect of query budget in ROUGE1 and ROUGE2 score.}
    \label{query}
\end{figure}

\textbf{Summary Evaluation.}
To avoid subjectivity in the evaluation process, we used the reference summaries as feedback.
The mentioned concepts that exist in reference summaries receive the maximum score by the ranked function.
We compared the summaries produced by three models, including the traditional approach (ExDos), a range of hierarchical approaches~\cite{christensen2014hierarchical}, and a structured summarization approach~\cite{falke2019automatic}, each tested on randomly selected documents from three datasets using ROUGE-1, ROUGE-2 and ROUGE-L scores based on the references summaries.
The average results reported in Table~\ref{tab:data1} show the supremacy of Summation in selecting specific contents.

\begin{table}[t]
\centering
  \caption{Comparing \textit{Summation} with benchmark datasets.}
  % \vspace{1mm}
  \label{tab:data1}
  \begin{tabular}{|l|c|c|c|}
    \hline
    Model &ROUGE1 &ROUGE2 &ROUGEL\\
    \hline
    Traditional Structured~\cite{falke2019automatic} &0.346  &0.090 &0.251\\
    \hline
    Traditional Hierarchical~\cite{christensen2014hierarchical}  &0.211  &0.013 &0.149\\
    \hline 
    \textit{Summation} &0.731 &0.651 &0.681\\
 \hline
\end{tabular}
\end{table}

\textbf{Query Budget Size.} We also measure the effectiveness of the users' query budget size in the process.
The pairwise preferences are defined based on the reference summaries, defining in a dictionary format.
We selected the query size among the selection of $\{10,15,20,25,30,35\}$, demonstrating the user's number of feedback.
The results are reported in Figure~\ref{query}.
As expected, by increasing the number of feedback, the ROUGE score increases significantly.
However, the difference rate decreases through the process.

% \begin{table}[t]
% \centering
%   \caption{Comparing SumRecom on DUC2002 dataset.}
%     \vspace{1mm}
%   \label{tab:data2}
%   \begin{tabular}{|l|c|c|c|}
%      \hline
%     Model &ROUGE1 &ROUGE2 &ROUGEL\\
%     \hline
%      APRIL &0.351 &0.078 &0.279\\
%       \hline
%      SPPI &0.350 &0.077 &0.278\\
%       \hline
%      SumRecom &0.35 &0.076 &0.279\\
%   \hline
% \end{tabular}
% \end{table}
% \begin{table}[t]
% \centering
%   \caption{Comparing SumRecom on DUC2004 dataset.}
%     \vspace{1mm}
%   \label{tab:data3}
%   \begin{tabular}{|l|c|c|c|}
%      \hline
%     Model & ROUGE1 & ROUGE2 & ROUGEL\\
%     \hline
%      APRI L&0.373 &0.093 &0.293\\
%       \hline
%      SPPI &0.372 &0.093 &0.293\\
%       \hline
%      SumRecom &0.382 &0.94 &0.301\\
%   \hline
% \end{tabular}
% \end{table}

\textbf{Human Analysis.}
Since the goal of \textit{Summation} is to help users make their desired summary, we conducted two human experiments to evaluate the model.
In the first experiments, to assess the possibility of finding their desired information, they were asked to answer a given question about each topic.
Their level of confidence in answering questions and their answers were recorded.
An evaluator assessed their accuracy in answering questions.
Among the fifteen workers, 86.67\% were completely confident in their answers.
However, 57\%  answered completely accurately.
In another task, after querying users for feedback, we ask them to select some concepts as the summary for the test data.
Then the outputs were also shown to users, and they all approved their satisfaction.
Besides, an evaluator manually compared them and reported more than 80\% correlation between outputs.

\section{Conclusion and Future Work}
\label{conclusion}
Extensive information in various formats is producing from single or multiple simultaneous sources in different systems and applications.
% such as medical records, customer databases, social networks, and business transaction systems.
% , results from scientific experiments, internet click-stream logs, real-time data sensor used in the internet of things (IoT) environments, or machine-generated data.
% Besides, data can be in various formats.
For instance, data can be structured, such as data in SQL databases, unstructured stored in NoSQL systems, semi-structured like web server logs, or streaming data from a sensor.
We propose a summarization approach based on a hierarchical concept map to tackle the variety and volume of big generated data.
% Moreover, the proposed approach interactively learns to generate personalized summaries based on users' feedback.
% We took a step further from the prior work by considering three research questions in this work:
% i) Can user preferences over concepts provide personalized summaries that reflect users' interests with less cognitive load?
% ii) Can domain-expert knowledge be embedded in the learning process?
% iii) How can user preferences and domain-experts experience be modeled as a reinforcement learning algorithm to generate desired summaries for users automatically?
We trained our approach using document collections as input and employed users' feedback to generate desired summaries for users, which can be extended to other data types.
Many future directions are possible.
First, capturing users' interests is a significant challenge in providing practical personalized information. 
The reason is that users are reluctant to specify their preferences as entering lists of interests may be a tedious and time-consuming process. 
% Besides, people's interests are not static and change over time that should be taken into account.
Therefore, techniques that extract implicit information about users' preferences are the next step for making useful personalized summaries.
Another potential direction is to use human feedback records to provide personalized summaries on new domains using transfer learning.
Moreover, we aim to use fuzzy clustering to make a hierarchical concept map.
% and use crowd knowledge for co-reference resolution.

% \textbf{\emph{Acknowledgement.}}We acknowledge the AI-enabled Processes (AIP\footnote{https://aip-research-center.github.io/}) Research Centre for funding this research.
% We also acknowledge Macquarie University for supporting this project through the IMQRES scholarship.
% \vspace{-4mm}
% \subsubsection{Brown Clustering} 
% Brown clustering [4], a popular class-based language model, can learn hierarchical clusters of words by maximizing the mutual information of word bigrams.
% Brown et al.[4] introduced a hierarchical clustering algorithm that maximizes the mutual information of wordbigrams.
% The probability for a set of words $w_1,w_2,...,w_T$ can be written as:
% \begin{equation}
%     \prod_{t=1}^T p(w_t|C(w_t))p(C(w_t)|C(w_{t−1}))
% \end{equation}
% where $C$ is a function that maps a word to its class, and $C(w_0)$ is a special start state.
% Brown clustering hierarchically merges clusters to maximize the quality of C.
% The quality is maximized when mutual information between all bigram classes are maximized. Although Brown clustering is commonly used, a major drawback is its limitation to learn only bigram statistics.
% We use an open-source implementation by Lianget al.[16].
% We have fine-tuned the number of cluster hyper-parameter by varying between 10 and 200.
%explain what is word embedding
% \bibliographystyle{elsarticle-num}

\section*{Acknowledgement}

We acknowledge the Centre for Applied Artificial Intelligence at Macquarie University, Sydney, Australia, for funding this research.

\bibliographystyle{IEEEtran}
\bibliography{ref.bib}

% \section*{References}

% \begin{thebibliography}{00}
% \bibliography{ref.bib}
% % \bibitem{b1} G. Eason, B. Noble, and I. N. Sneddon, ``On certain integrals of Lipschitz-Hankel type involving products of Bessel functions,'' Phil. Trans. Roy. Soc. London, vol. A247, pp. 529--551, April 1955.
 
% \end{thebibliography}

\end{document}